\documentclass[11pt]{article}

\usepackage{graphicx,epsfig}
\usepackage{multicol}
\usepackage{amsmath,amssymb,cite,color,hyperref}
\begin{document}

\begin{flushright}
CERN-PH-TH/2013-204\\
\end{flushright}

\begin{center}
{\large\bf Bayesian reweighting of nuclear PDFs\\ and constraints from proton-lead collisions at the LHC}
\vspace{0.6cm} \\

Nestor Armesto,$^{1,2}$ Juan Rojo,$^2$
Carlos A. Salgado,$^1$ and Pia~Zurita.$^1$

\vspace{.3cm}
{\it ~$^1$ Departamento de F\'isica de Part\'iculas and \\ IGFAE, Universidade de Santiago de Compostela,\\
E-15706 Santiago de Compostela, Galicia, Spain \\
~$^2$ PH Department, TH Unit, CERN, CH-1211 Geneva 23, Switzerland \\}
\end{center}   

\vspace{0.2cm}

\begin{center}
{\bf \large Abstract}
\end{center}
New hard-scattering measurements from the LHC proton-lead run have the potential to provide important constraints on the nuclear parton distributions and thus contributing to a better understanding of the initial state in heavy ion collisions.
 In order to quantify these constraints, as well as to assess the compatibility with available nuclear data from fixed target experiments and from RHIC, the traditional strategy is to perform a global fit of nuclear PDFs.
 This procedure is however time consuming and technically challenging, and moreover can only be
performed by the PDF fitters themselves.
 In the case of proton PDFs, an alternative approach has been suggested that uses Bayesian inference to propagate the effects of new data into the PDFs without the need of refitting.
 In this work, we apply this reweighting procedure to study the impact on nuclear PDFs of low-mass Drell-Yan and single-inclusive hadroproduction pseudo-data from proton-lead collisions at the LHC as representative examples.
 In the hadroproduction case, in addition we assess the possibility
of discriminating between the DGLAP and CGC production frameworks.
 We find that LHC proton-lead data could lead to a substantial reduction of the uncertainties on nuclear PDFs, in particular for the small-$x$ gluon PDFs where uncertainties could decrease by up to
a factor two.
The Monte Carlo replicas of EPS09 used in the analysis are released as a public code for general use. It can be directly used, in particular, by the experimental collaborations to check, in a straightforward manner, the degree of compatibility of the new data with the global nPDF analyses.

\clearpage

\tableofcontents

\section{Introduction}

The knowledge of the parton distribution functions  (PDFs) of the proton has substantially improved in the last years~\cite{Ball:2012cx,Martin:2009iq,Gao:2013xoa,Ball:2012wy,Forte:2013wc}, thanks to the  
increased coverage and variety of experimental data
included,  theoretical improvements in higher order computations,
 as well as from methodological developements. 
Despite recent developements~\cite{Eskola:2009uj,Helenius:2012wd,deFlorian:2011fp,Hirai:2007sx,Schienbein:2009kk}, the determination of nuclear PDFs (nPDFs) has not reached the same level of accuracy, due to the scarce amount of available data and the limited kinematical coverage of the measurements done up to date --- see e.g. the review \cite{Armesto:2006ph}). 
In this respect, measurements from the recent LHC proton-lead run have the potential~\cite{Salgado:2011wc,Albacete:2013ei,Paukkunen:2010qg,Stavreva:2010mw,Arleo:2011gc,Kang:2012am,Eskola:2013aya} to be of great relevance as they will not only extend the probed kinematical range but also provide information on heavier nuclei where data is currently very limited.
In turn, more accurate nPDFs allow for a more reliable characterization of the initial state in heavy ion collisions.

In order to extract information out of a novel measurement, a new determination of PDFs must be done, by including the  new data in a global fit with all preceding data. 
Though straightforward in principle, performing a new fit is a cumbersome and time-consuming process. 
In addition, exploring formerly unknown regions (such as small-$x$) might required substantial modifications on the theoretical input and the fitting parameterization strategy. 
That is, it is not clear just by looking at the new data whether it is compatible or not with previous results.
 This is particularly true in the case of experiments involving nuclei, as the nuclear medium might present a variety of more complex phenomena than a simple modification of the parton distributions.

As an alternative to repeating the nPDF global fit,
 it is possible to reweight an existing
PDF set with the information contained in the new measurement using Bayesian inference, applying the techniques originally developed
 for proton PDFs~\cite{Ball:2010gb,Ball:2011gg}.
 This method allows to study quantitatively both the compatibility of new data with that used in the original PDFs determination and to determine its impact on the
central values and uncertainbties of the PDFs.
 While the original derivation~\cite{Ball:2010gb,Ball:2011gg} applied only to PDF sets based on the Monte Carlo method, it was later shown~\cite{Watt:2012tq} how the same method can be extended to Hessian PDFs, which is the framework adopted for most nuclear PDF sets available.

The aim of this work is thus to perform an exploratory quantitative study of the constraining potential for nuclear PDFs of the LHC proton-lead run data.
We have selected two representative processes: low-mass Drell-Yan production and
charged hadron inclusive production.
Since no data from $p$Pb collisions in the hard scattering regime is yet available (except for charged particle production in the pilot run from ALICE~\cite{ALICE:2012mj}), we will simulate pseudo-data based on a known underlying theory, namely the collinear DGLAP framework and the {\tt EPS09} nuclear PDF set.
In the case of hadroproduction, in addition pseudo-data has been also simulated in the Color Glass Condensate (CGC) 
framework (see e.g. the review~\cite{Albacete:2013tpa}): 
this allows to quantify to
which extent non-linear effects in charged hadron production can be absorbed
in a global nPDF fit based on the DGLAP framework,  
While our analysis is based on the {\tt EPS09} nuclear PDF set, the qualitative
results should be valid for all other nPDF sets.

As an extra, we also release the set of Monte Carlo replicas of EPS09 as a public computer code\footnote{The code can be downloaded from \url{http://igfae.usc.es/hotlhc/index.php/software}}. These replicas can be used directly, in particular by the experimental collaborations, to check the compatibility of the new data with the nuclear parton distributions, as well as to pindown the corresponding constraints for each parton flavor. The procedure to do that is the same as applied here.

This article is organized as follows: first of all, 
in Sec.~\ref{sec:weight} we present a brief summary of the main features of the reweighting procedure and construct a Monte Carlo version of the {\tt EPS09} set.
Then we study the impact of Drell-Yan production in {\tt EPS09} in Sec.~\ref{sec:dy}, 
before moving to
Sec.~\ref{sec:charged} where we study charged hadron production.
In this latter case, we explore both the constraints on nuclear PDFs and
the discrimination power between DGLAP and CGC scenarios.
Sec.~\ref{sec:summary} summarizes our results and discusses the prospects
for other relevant measurements.

\section{Bayesian reweighting of nuclear PDFs \label{sec:weight}}

PDF uncertainties can be determined using basically two methods: the Hessian
approach (with and without tolerance), upon which all nuclear PDF sets are based, and the Monte Carlo approach.
In a Monte Carlo PDF set, such as those of the  NNPDF Collaboration~\cite{Ball:2008by,Ball:2010de}, the underlying
PDF probability density  $\mathcal{P}$ is sampled by generating, through a Monte Carlo procedure, an ensemble of $N_{\rm rep}$ PDFs replicas $f_{k},\,k=1,..,N_{\rm rep}$, each fitted to a replica of the experimental data.
Then any quantity $\mathcal{O}[f]$ depending on the PDFs can be evaluated by computing $\mathcal{O}[f_k]$ with $k=1,...,N_{\rm rep}$ and averaging over the results for individual replicas,
\begin{equation}
\langle \mathcal{O} \rangle=\frac{1}{N_{\rm rep}}\sum_{k=1}^{N_{\rm rep}}\mathcal{O}[f_{k}] \, .
\label{eq:avgold}
\end{equation}

Consider now a new measurement consisting of \emph{n} points with
 covariance matrix ${\rm cov}_{ij}$, not included in the orginal determination of
 $\mathcal{P}$,
\begin{equation}
y=\lbrace y_{1}, y_{2}, ..., y_{n} \rbrace \, .
\label{eq:data}
\end{equation}  
Using Bayesian inference~\cite{Ball:2010gb,Ball:2011gg}, it is possible to update the original probability distribution $\mathcal{P}_{\rm old}(f)$ to a new probability distribution $\mathcal{P}_{\rm new}(f)$ that accounts for the information contained in the new
measurement.
This can be achieved by computing the new
 weight $w_{k}$ for each individual replica $f_{k}$, which measures its agreement with the new data. 
It should be noted that, as shown explicitely in Ref.~\cite{Ball:2010gb},
Bayesian reweighting is fully equivalent to a full new fit, provided the new
data is not too constraining so that the effective number of replicas
is still large enough (see below).

Following~\cite{Ball:2010gb,Ball:2011gg},  these weights turn out to be
\begin{equation}
w_{k}=\frac{(\chi_{k}^{2})^{\frac{1}{2}(n-1)}e^{-\chi_{k}^{2}/2}}{\frac{1}{N_{\rm rep}}\sum_{k=1}^{N_{\rm rep}}(\chi_{k}^{2})^{\frac{1}{2}(n-1)}e^{-\chi_{k}^{2}   /2}} \, ,
\label{eq:w}
\end{equation}
in terms of the $\chi^2$ for each replica between the original theory predictions for the $k$-th replica and the new experimental
measurement,
\begin{equation}
\chi^{2}_{k}(y,f_{k})=\sum_{i,j=1}^{n}(y_{i}-y_{i}[f_{k}]){\rm cov}_{ij}^{-1}(y_{j}-y_{j}[f_{k}]) \, ,
\label{eq:chi2}
\end{equation} 
so that now the analog of Eq.~(\ref{eq:avgold}) reads
\begin{equation}
\langle \mathcal{O} \rangle_{\rm new}=\frac{1}{N_{\rm rep}}\sum_{k=1}^{N_{\rm rep}}w_{k}\mathcal{O}[f_{k}] \, .
\label{eq:avgnew}
\end{equation} 

The only feature that distinguishes between a full new fit from the Bayesian reweighting is the statistical efficiency of the latter.
 While a new fit would give the best representation of the underlying density for a given $N_{\rm rep}$, this is not the case for the PDF reweighting. The replicas with very small weights will become almost irrelevant when computing averages and the accuracy of the representation of the underlying distribution $\mathcal{P}_{\rm new}(f)$ will decrease.
 To quantify this efficiency loss, the Shannon entropy can be used to compute $N_{\rm eff}$, the effective number of replicas after reweighting:
\begin{equation}
N_{\rm eff}\equiv \exp\left\{ \frac{1}{N_{\rm rep}}\sum_{k=1}^{N_{\rm rep}}w_{k}\log (N_{\rm rep}/w_{k})\right\} \, .
\label{eq:shannon}
\end{equation}
The above equation determines that the accuracy of the reweighted fit is the same that would be obtained if a new fit with $N_{\rm eff}$ replicas were to be performed.
When $N_{\rm eff} \ll N_{\rm rep}$ the reweighting method becomes unreliable,
and a full refit is mandatory.
This scenario might arise either because the new data is inconsistent with the 
original one within the given theoretical framework, or when the new data contains substantial new
 information on the PDFs as
compared to that in the original determination.

Two consistency tests can be performed in order to clarify this issue. The first one is the examination of the $\chi^{2}$ profile of the new data,
\begin{equation}
\mathcal{P}(\chi^{2})=\frac{1}{N_{\rm rep}}\sum_{k=1}^{N_{\rm rep}}w_{k} \, ,
\label{eq:chi2profile}
\end{equation}
with $k$ all replicas such that $\chi^{2}_{k}\in[\chi^{2},\chi^{2}+d\chi^{2}]$. 
Alternatively, it is possible to estimate if the agreement between
data and theory improves if we rescale the experimental uncertainties by a factor $\alpha$, then the probability density for the rescaling parameter will be
\begin{equation}
\mathcal{P}(\alpha)\varpropto\frac{1}{\alpha}\sum_{k=1}^{N_{\rm rep}}w_{k}(\alpha) \, ,
\label{eq:alphaprofile}
\end{equation}
where $w_{k}(\alpha)$ are the weights of Eq.~(\ref{eq:w}) evaluated by replacing $\chi^{2}_{k}$ by $\chi^{2}_{k}/\alpha^{2}$ (proportional to the probability of $f_{k}$ given a data set with rescaled errors). 
When the $\mathcal{P}(\alpha)$ is peaked around one, the new data are consistent with the initial theory distribution.
If $\mathcal{P}(\alpha)$ is peaked at a value larger than one, this 
might suggest either that experimental uncertainties have  been underestimated
or that the theory framework that is used is not the right one to
describe this measurement.

While the derivation above that leads to the
weights for each replica Eq.~(\ref{eq:w}) 
applies to PDF sets based on the Monte Carlo method, the goal
of this paper is to study the impact on nuclear PDFs, for which all
sets with uncertainty bands available are based on the Hessian method.
As discussed in Ref.~\cite{Watt:2012tq}, it is possible to generate
Monte Carlo replicas starting from Hessian PDF error sets, and then
apply the standard formulae such as Eq.~(\ref{eq:avgnew}).
We have thus constructed a Monte Carlo version of the  {\tt EPS09}~\cite{Eskola:2009uj} nuclear PDF set, as follows.
For each parton flavor we took the central PDF set ($f_{0}$) and shifted it using the Hessian error sets according to
\begin{equation}
f_{k}(x,Q^2)=f_{0}(x,Q^2)+\sum_{i}^{N_{\rm eig}}\left(f_{i}^{\pm}(x,Q^2)-f_{0}(x,Q^2)
\right)\lvert r_{k,i}\rvert \, , \quad (k=1,\ldots,N_{\rm rep})\, , 
\label{eq:repgen}
\end{equation}
where $N_{\rm eig}$ is the number of pairs of Hessian eigenvectors, $N_{\rm eig}=15$ in the particular case of {\tt EPS09}, and  $r_{k,i}$ are random numbers from a Gaussian distribution of mean zero and variance one, and $f^{\pm}_{i}$ is the nPDF
corresponding to the eigenvector $S^{\pm}_{i}$, and we use the positive sign for $r_{k,i}>0$ and the negative sign for $r_{k,i}<0$. 
Note that a symmetric version of Eq.~(\ref{eq:repgen}) can be obtained by averaging over each pair of eigenvectors~\cite{Watt:2012tq}.

\begin{figure}[t]
\begin{center}
\vspace*{-0.6cm}
\epsfig{figure=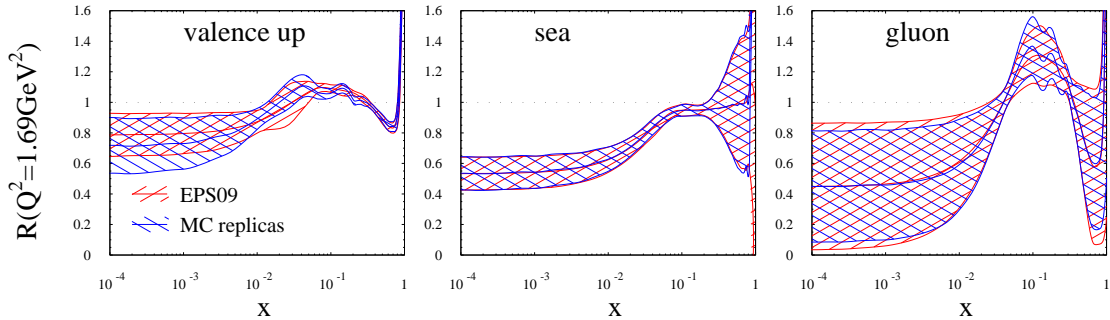,width=1.00\textwidth}
\end{center}
\vspace*{-0.5cm}
\caption{\label{fig:check}
Comparison of the nuclear ratios $R_i^A(x,Q^2)$  for lead 
($A=208$) for three flavours at $Q^{2}=1.69$ GeV$^{2}$ in {\tt EPS09}, comparing the original Hessian PDFs with their
Monte Carlo counterparts.
From left to right: up valence, quark sea, and gluon nuclear ratios.}
\end{figure}

Following Eq.~(\ref{eq:repgen}), we have generated 
 $N_{rep}=1000$ Monte Carlo replicas for {\tt EPS09}. 
We have checked the consistency of the procedure for all
nuclear PDFs, by comparing central values
and 1-sigma uncertainties in the Hessian and MC versions of {\tt EPS09}, and
finding reasonable agreement within the statistical accuracy expected.
As an illustration, in Fig.~\ref{fig:check} we show the nuclear ratios for lead ($A=208$), defined
as the ratio of nuclear PDFs for nuclear number $A$ divided by the corresponding proton PDFs,
\begin{equation}
R_i^A(x,Q^2)=\frac{f_i^A(x,Q^2)}{f_i^{p}(x,Q^2)}
\end{equation}
 at $Q^{2}=1.69$~GeV$^{2}$ for the up valence  (left), quark sea (middle) and gluon (right) nuclear PDFs.\footnote{Note that the wiggles that can be observed in the Monte Carlo version of {\tt EPS09} for the gluon and the up valence quark near $x\sim 0.1$ in Fig.~\ref{fig:check}  arise from the interpolation in the {\tt EPS09} driver routines.
 These wiggles disappear if the symmetrised version of Eq.~(\ref{eq:repgen}) is used. They have a negligible impact on our results, since the LHC $p$Pb data
that we consider affects smaller values of $x$.}
As we can see, both mean values and 1-sigma uncertainties  are in reasonable agreement between the Hessian and Monte Carlo versions of {\tt EPS09}.
The remaining differences can be explained since we are neglecting (small)
corrections beyond the linear approximation which is always assumed
in Hessian PDF sets~\cite{Watt:2012tq}.

Let us also mention that Ref.~\cite{Paukkunen:2013grz} proposes an alternative approach for including new data into a nuclear PDF fit similar
to the reweighting method just discussed.

\section{Constraints from Drell-Yan production\label{sec:dy}}

After describing our framework, we 
start by studying the impact on nuclear PDFs of  Drell-Yan production
 in proton-lead collisions at the LHC.
In proton-proton collisions, neutral current Drell-Yan production
provides important constraints on the proton PDFs, in particular for
quarks and antiquarks, but also for gluons,
 and has been measured at the LHC by ATLAS, CMS and
LHCb~\cite{Chatrchyan:2011cm,Aad:2013iua,lhcb}.
On the other hand, 
the amount of Drell-Yan data so far included in nuclear PDFs fits is 
 fairly small 
and affected by substantial experimental uncertainties, so the constraints derived from it are not as tight as the ones on nuclear valence quark distributions from 
deep-inelastic scattering (DIS)
data.

In this respect, Drell-Yan data from proton-lead collisions at the 
LHC should  provide crucial information to further constrain the gluon density at small and medium $x$ as well as the sea quark distributions.
We are in particular interested in low-mass Drell-Yan production, since
on the one hand the small scale implies enhanced sensitivity to nuclear effects  and in addition the (off-peak) cross-section rises
when the invariant mass of the Drell-Yan pair $m_{ll}$ decreases.
On the other hand, the small $m_{ll}$ regime is also interesting since some
 studies~\cite{GolecBiernat:2010de} 
predict that non-linear effects might show up as a substantial difference
as compared to the linear DGLAP framework in which nuclear PDF fits
are based.

 Numerical simulations for DY production in $p$Pb collisions applying the same kinematics cuts that were used in the $pp$ case show that, given the integrated luminosity of the present run, around $\mathcal{L}_{\rm int}\sim 30$ nb$^{-1}$, very few events would be obtained in the forward region, as well as in the central region if the same cuts on lepton $p_{PT,l}$ were to be imposed.
 Therefore, for the purposes of this exercise we choose to consider only Drell-Yan cross-sections for $\lvert \eta \rvert<4$, without any kinematical cut on $p_{T,l}$. 
Of course, the actual cuts will be different in real analysis, but for
instance the cut in $p_{T,l}$ can be reduced as compared to $pp$ since
there are no problems with triggering most events.

To generate both the pseudo-data and the predictions for DY production based on the $N_{\rm rep}=1000$ replicas of the MC version of {\tt EPS09}, 
we calculate the cross-sections in the low-mass range $m_{ll}<12$ GeV by use of the {\tt MCFM} code~\cite{MCFM}, modified in order to account for the fact that one of the initial particles is a lead nucleus.
 The pseudo-data was computed from the central values of {\tt EPS09} and adding
the corresponding statistical fluctuations.
The statistical uncertainties were computed from the number of
expected events in each bin, and a 
total uncorrelated 8\% systematic uncertainty was also assumed. 
This value is a conservative estimate, based on the result that
 in the proton-proton case the systematic uncertainty is about 4\% for the $15$ GeV$<m_{ll}<20$ GeV invariant mass bin~\cite{Chatrchyan:2011cm}.
For the proton PDFs we used MSTW08~\cite{Martin:2009iq}, though results were
essentially unchanged if some other proton PDF set was used.
In the following, PDF uncertainties arise only from the {\tt EPS09} nuclear
PDFs, neglecting the proton PDFs.

\begin{figure}[t]
\begin{center}
\vspace*{-0.6cm}
\epsfig{figure=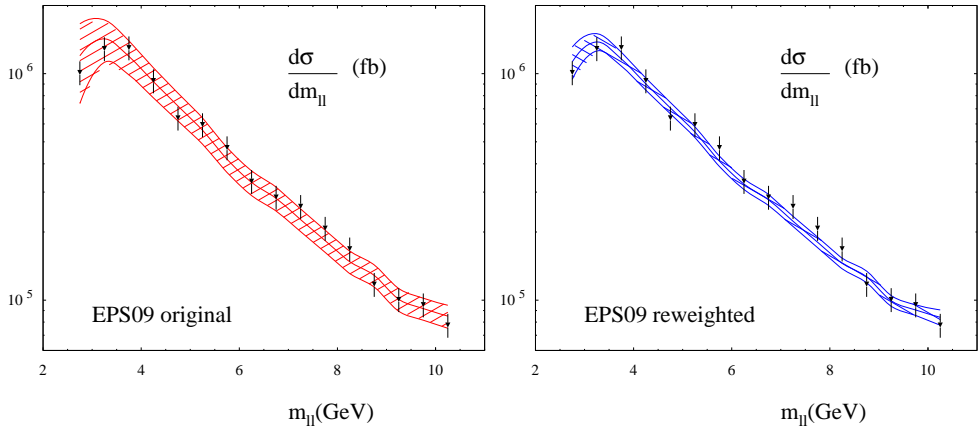,width=0.90\textwidth}
\end{center}
\vspace*{-0.5cm}
\caption{\label{fig:dy} The predictions for the Drell-Yan differential
cross sections with {\tt EPS09} compared to pseudo-data as a function of the invariant lepton pair mass $m_{ll}$ at central rapidity before (left) and after (right) PDF reweighting.
The error band corresponds the 1-sigma uncertainty in {\tt EPS09}.
 }
\end{figure}

To begin with, in Fig.~\ref{fig:dy} we show the pseudodata and the predictions before (left) and after (right) reweighting. The error band in the prediction accounts only for the  uncertainties of the {\tt EPS09} nuclear PDFs. 
It is clear that once the new data is included into the nuclear PDFs, the uncertainty band of the theory prediction is reduced, without affecting the central value, as expected for perfectly consistent data.
 In Table~\ref{tab:chi2} we provide the effective number of replicas $N_{\rm eff}$, together with the values for the mean $\chi^{2}$ and the average over replicas $\langle \chi^{2}\rangle$ per data point before and after the reweighting. 
%

\begin{table}[h]
\begin{centering}
\begin{tabular}{cccc}
\hline
   & $\chi^{2}/n$ & $\langle \chi^{2}\rangle/n$    & $N_{\rm eff}$   \\ \hline
\hline
 Original & 0.64 & 2.68 &   -  \\
 Reweighted & 0.59 & 0.96   &   539\\  \hline
\end{tabular}
\caption{\label{tab:chi2} $\chi^{2}/n$ and $\langle \chi^{2}\rangle/n$ values before and after the reweighting of EPS with the Drell-Yan pPb pseudo-data, with $n=16$ points. 
The effective number of replicas $N_{\rm eff}$ is also provided for the reweighting case.}
\end{centering}
\end{table}

From Table~\ref{tab:chi2} we see that the $\chi^{2}/n$ is $\mathcal{O}(1)$, as
expected by the use of consistent pseudo-data, while $\langle \chi^{2} \rangle /n\sim 2.7$, indicating that some replicas are clearly disfavoured by the pseudo-data.
 After the reweighting  the weight for replicas far from the pseudo-data is suppressed, leadind to a substantial reduction in $\langle \chi^{2}\rangle/n$:
the reweighted sample includes only replicas that agree with the 
pseudo-data.
The number of effective replicas for this pseudo-data set is $N_{\rm eff}$=539, while the original sample had $N_{\rm eff}$=1000, showing that is about half of the replicas are strongly disfavoured after the inclusion of the Drell-Yan pseudo-data.

\begin{figure}[t]
\begin{center}
\vspace*{-0.6cm}
\epsfig{figure=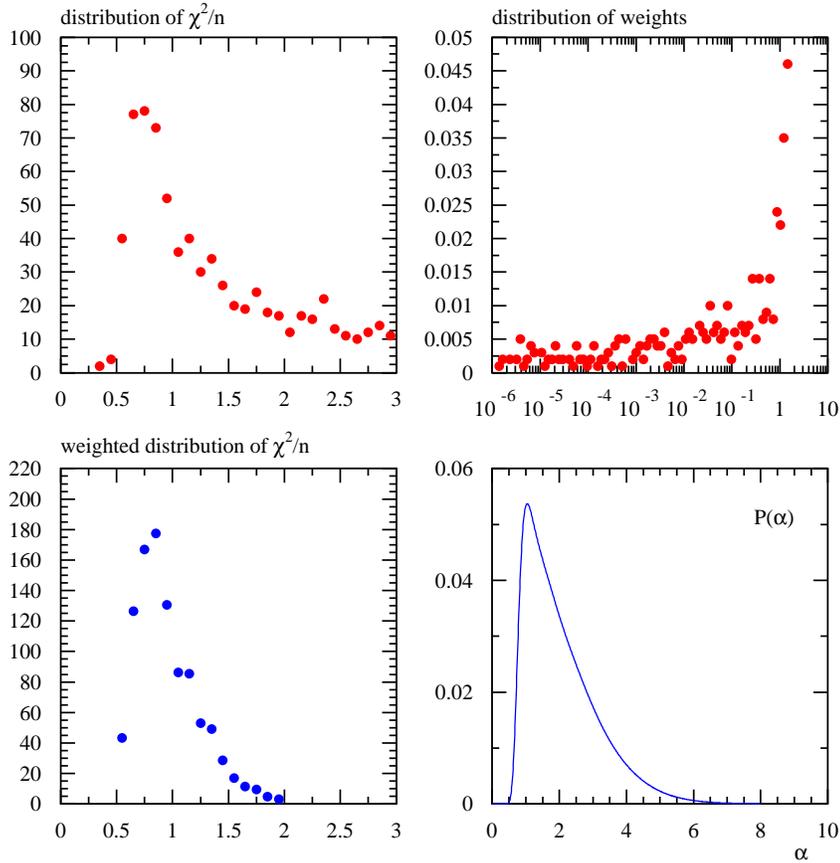,width=0.85\textwidth}
\end{center}
\vspace*{-0.5cm}
\caption{\label{fig:dy-dist}
Left plots: distribution of the $\chi^{2}_{k}/n$ ($\chi^{2}$ per data point) before (upper) and after (lower) the reweighting of {\tt EPS09} with pseudo-data for Drell-Yan production in $p$Pb collisions. 
Upper right plot: distribution of the weights $w_{k}$. Lower right plot: $\mathcal{P}(\alpha)$.}
\end{figure}

Then we apply the consistency tests as described in Sect.~\ref{sec:weight} and present the $\chi^{2}$, $w_{k}$ and $\mathcal{P}(\alpha)$ density distributions in Fig.~\ref{fig:dy-dist}. 
The $\chi^{2}$ density distribution before the reweighting (upper left) is peaked around $0.8$ but with a tail towards higher $\chi^{2}$ values. Then, as we already knew, the pseudo-data is compatible with the one used in {\tt EPS09} fit and its inclusion in a refit should have a moderate impact.
 We confirm this by looking the $\chi^{2}$ after the reweighting (lower left): the peak shifts towards 1 and the tail is significantly reduced.
 One last check in that regard is the $\mathcal{P}(\alpha)$ distribution in the lower right plot. The most probable value for the error rescaling parameter $\alpha$ is almost 1, so our error estimation was quite good.

\begin{figure}[t]
\begin{center}
\vspace*{-0.6cm}
\epsfig{figure=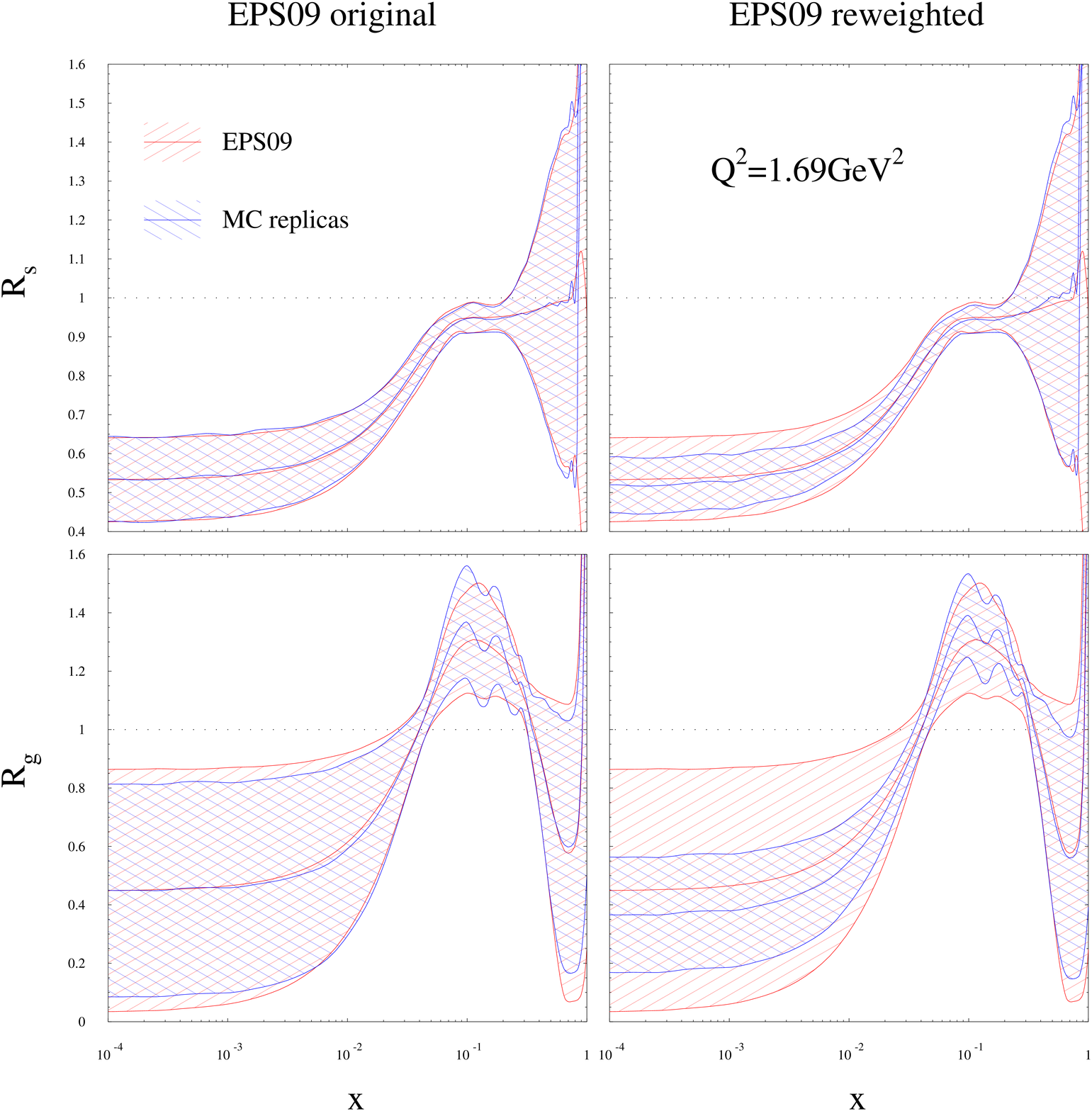,width=0.85\textwidth}
\end{center}
\vspace*{-0.5cm}
\caption{\label{fig:dy-sg}
{\tt EPS09} nuclear ratios $R^A_i(x,Q^2)$ for  $Q^{2}=1.69$~GeV$^{2}$, both
in the Hessian and in the Monte Carlo versions,  before
(left plots) and after (right plots) the reweighting with LHC $p$Pb Drell-Yan production
pseudo-data. We show both the nuclear modifications of sea quarks (upper plots)
and of gluons (lower plots). }
\end{figure}

We can now assess the impact of the DY pseudo-data on the nuclear PDFs.
 In this case while for the up and down valence distributions there seems to be no significant change (plot not shown), the sea distributions present a reduction on the respective uncertainty bands in the $x<10^{-2}$ region accompanied of a very slight but negligible decrease of the central value.
 This is illustrated in the upper plots of Fig.~\ref{fig:dy-sg}, where we show the total sea quark nuclear ratio. 
The same behaviour is found for all sea distributions. The one parton that varies distinctly is however the gluon, as can be seen in the 
lower plots of Fig.~\ref{fig:dy-sg}. 
When $x$ is below $10^{-2}$ there is a displacement of the central curve towards lower values accompanied by a reduction of the uncertainty of about $50\%$: the pseudo-data seems to favor more suppressed gluons in that region.

To summarize, the analysis of Drell-Yan pseudo-data 
from $p$Pb collisions shows that (assuming full compatibility with
the collinear framework) this measurement has a strong potential
and would provide useful information to improve our knowledge of the gluon density in lead nuclei, which is of course a crucial ingredient to characterize
the initial state in nuclear collisions at the LHC.

\section{Constraints from inclusive hadron production\label{sec:charged}}

Now we turn to the study of potential constraints on nPDFs
arising from  charged hadron single inclusive production, defined
as the sum over all final state charged mesons and baryons.
 In this case our observable will be the cross section in proton-lead collisions divided by the same quantity in proton-proton collisions. 
This is the way in which  experimental collaborations typically present their results, since the ratio cancels several experimental
uncertainties and on top removes part of the proton PDF dependence.

 We will consider two scenarios: one in which our pseudo-data is generated from CGC predictions, which include non-linear effects, and another using the collinear DGLAP framework and {\tt EPS09}, which assume linear QCD evolution.
The analysis based on CGC pseudo-data
 is relevant to determine the discriminating power of this measurement 
regarding saturation dynamics. 
It should be taken into account
 that at least part of possible non-linear 
effects, if present in data, can be
absorbed in the DGLAP fit, as discussed in Refs.~\cite{Caola:2009iy,Albacete:2012rx} regarding this same problem in
 electron-proton collisions.
On the other hand, when the pseudo-data is generated using {\tt EPS09},
 we can determine the
improvement in the accuracy of nuclear PDFs in the case of a consistent
underlying theory.

Let us mention that inclusive charged particle production
is closely related to inclusive pion production, 
another process that has been used in the past to constrain nPDFs
using $d$Au collisions from RHIC, although
 available data from RHIC is scarce and affected
by large uncertainties.
 While it would be more realistic
(from the experimental point of view)
 to include in our analysis pseudo-data for identified pions and 
kaons rather than sum inclusively over all 
charged particles, we have done so in order to compare with the available
 CGC predictions.

\subsection{Hadroproduction in the DGLAP framework \label{sec:chargedeps}}

Let us begin with the analysis of inclusive charged hadron production
 using pseudo-data generated with {\tt EPS09} central values in the collinear DGLAP framework, as done
in the previous section for the DY process.
The statistical uncertainties are determined from the expected
number of events in each data bin, and we have assumed 5\% and 7\% (uncorrelated) systematic and normalization uncertainties respectively, slightly larger than the
corresponding proton-proton results~\cite{Khachatryan:2010us,Aad:2010ac,Abelev:2013ala}.
Theoretical predictions have been computed using the code
for NLO inclusive hadron production of Ref.~\cite{Jager:2002xm}, modified
to account for nuclear effects as discussed in Ref.~\cite{Sassot:2010bh}.
 No nuclear effects were taken into account for the fragmentation in the nuclear medium, and the 
DSS~\cite{de Florian:2007hc} fragmentation functions 
(FFs) were employed. 
%

\begin{figure}[h]
\begin{center}
\vspace*{-0.6cm}
\epsfig{figure=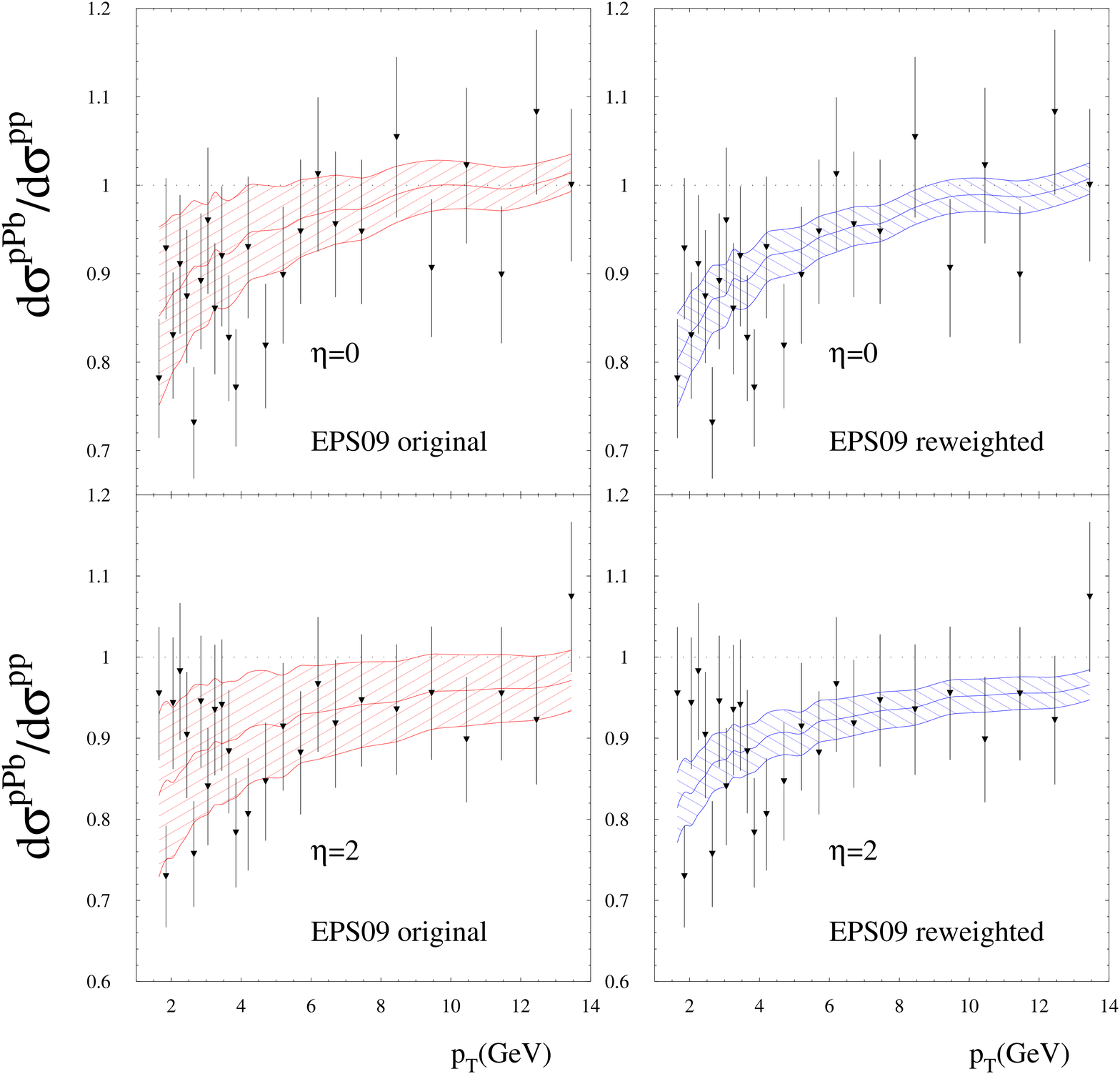,width=0.8\textwidth}
\end{center}
\vspace*{-0.5cm}
\caption{\label{fig:eta-eps}
Upper plots: ratio of the charged hadron single inclusive cross section in $p$Pb with respect to $pp$ collisions, as a function of the hadron 
transverse momentum $p_{T}$ at central rapidity before (left) and 
after (right) the reweighting, where the pseudo-data has been generated
in the collinear DGLAP framework
using {\tt EPS09}.
Lower plots: the same for forward rapidities, $\eta=2$.}
\end{figure}

In Fig.~\ref{fig:eta-eps} we present the comparison between the
 the cross sections for the pseudo-data and for the {\tt EPS09} predictions, for
two different hadron rapidities, central ($\eta=0$) and
forward ($\eta=2$).
 After including the pseudo-data into the nuclear fit by reweighting,
we find a shift of the central values, suppression for $p_{T}<8$ GeV and enhancement above, 
together with  a clear narrowing of the uncertainties. 
Then in Table~\ref{tab:chi2-eps} we show the values of $\chi^{2}/n$, $\langle \chi^{2}\rangle/n$ and $N_{\rm eff}$. For both rapidities almost two thirds of the replicas survive the reweighting. 
The improvement in $\langle \chi^{2}\rangle/n$ after including the pseudo-data indicates, as
for DY, that theoretical predictions inconsistent with the pseudo-data have been effectively
removed.

\begin{table}[h]
\begin{center}
\begin{tabular}{cccccccccc}
\hline
 & & $\eta=0$ & & \vline& & $\eta=2$     \\ \hline
 & $\chi^{2}/n$ & $\langle \chi^{2}\rangle/n$ & $N_{\rm eff}$ & \vline & $\chi^{2}/n$ & $\langle \chi^{2}\rangle/n$ & $N_{\rm eff}$  \\ \hline\hline
 Before & 1.11 & 1.75 & & \vline & 0.95 & 1.82 &      \\
 After & 0.84 & 1.02 & 624 & \vline& 0.92 & 1.08 & 612     \\  
\hline
\end{tabular}
\caption{\label{tab:chi2-eps} Same as Table~\ref{tab:chi2} for
the reweighting of {\tt EPS09} with inclusive charged hadron production data, for central ($\eta=0$) and forward ($\eta=2$) rapidities.
Pseudo-data has been generated in the DGLAP framework.
}
\end{center}
\end{table}

\begin{figure}[h]
\begin{center}
\vspace*{-0.6cm}
\epsfig{figure=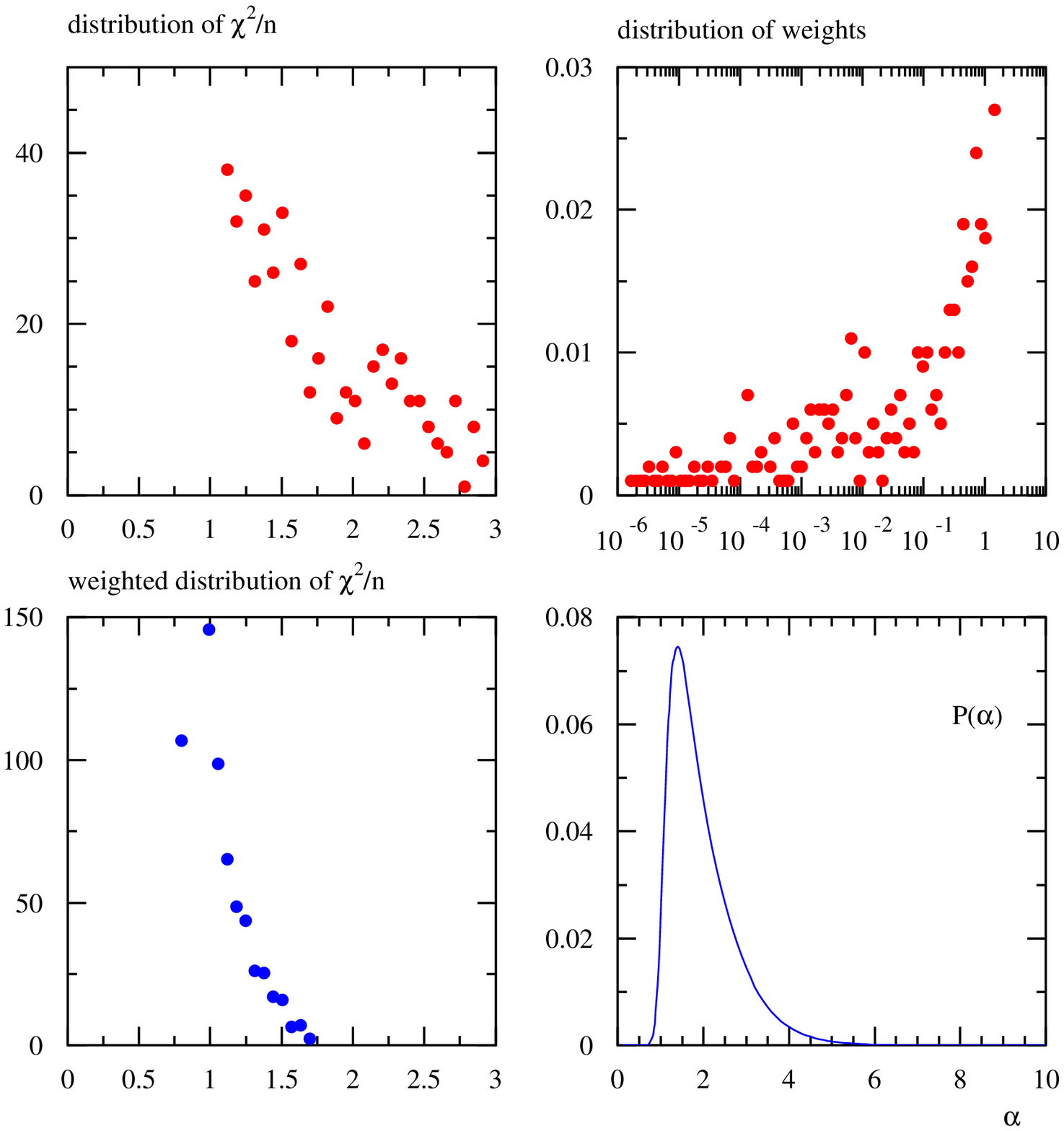,width=0.85\textwidth}
\end{center}
\vspace*{-0.5cm}
\caption{\label{fig:eta-dist-eps}
Same as Fig.~\ref{fig:dy-dist} for inclusive charged hadron central production
$\eta=0$, with pseudo-data generated in the DGLAP framework with {\tt EPS09}. }
\end{figure}

We then consider the consistency tests and in Fig.~\ref{fig:eta-dist-eps} show the $\chi^{2}$, $w_{k}$ and $\mathcal{P}(\alpha)$  distributions in the central region ($\eta=0$). Similar information is
derived from the analysis of the forward region.
Before the reweighting the $\chi^{2}$-density distribution (upper left) is peaked around $1.25$, with a tail towards higher $\chi^{2}$; afterwards (lower left) the peak moves towards one and the tail drastically reduces. 

\begin{figure}[t]
\begin{center}
\vspace*{-0.6cm}
\epsfig{figure=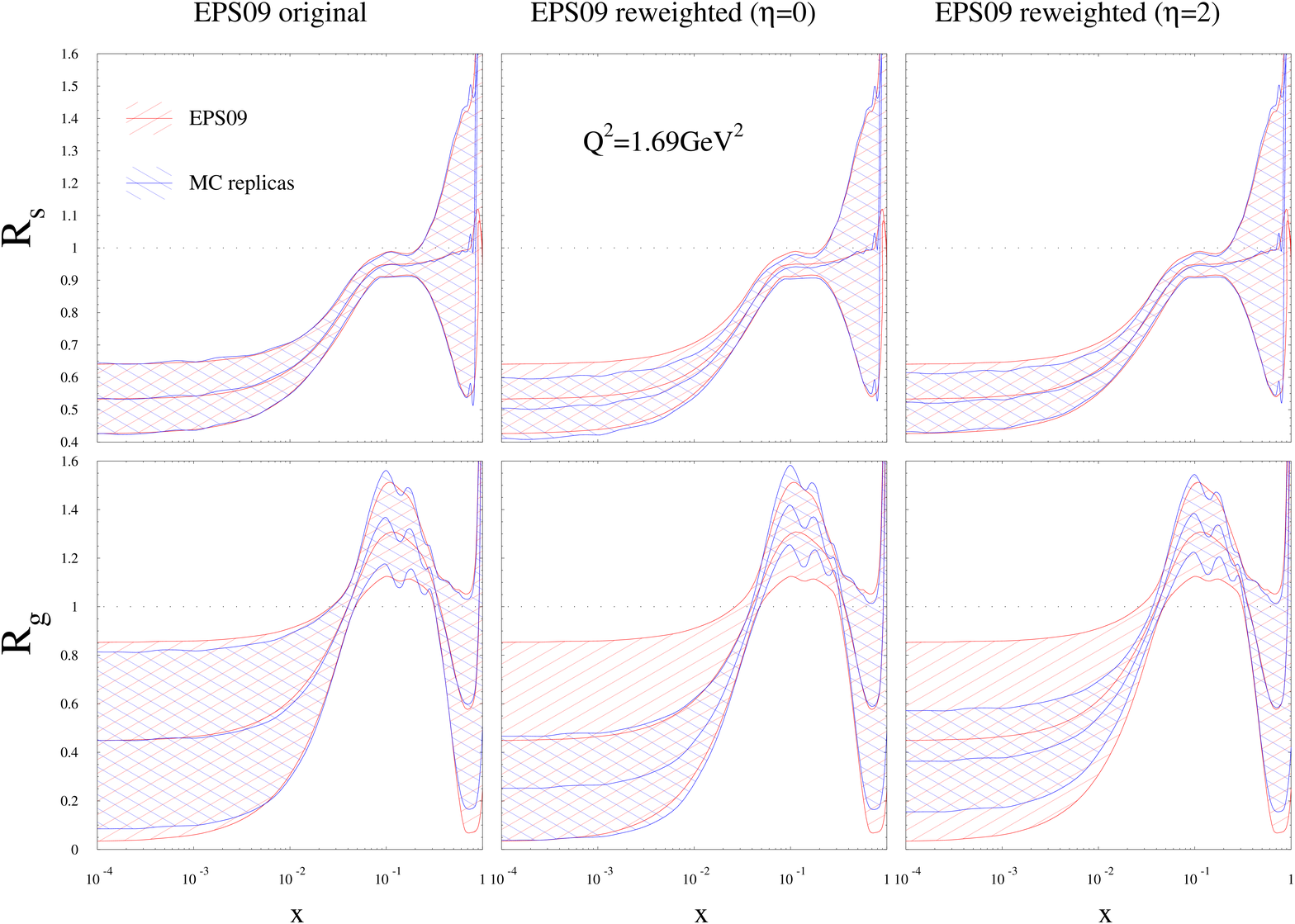,width=1.00\textwidth}
\end{center}
\vspace*{-0.5cm}
\caption{\label{fig:eta-eps-w}
Same as Fig.~\ref{fig:dy-sg}, now for the nuclear modifications of sea quarks (upper plots)
and gluons (lower plots) when $p$Pb LHC pseudo-data for inclusive charged hadron production
at central ($\eta=0$) and forward ($\eta=2$) rapidities is included in {\tt EPS09}. The
pseudo-data has been generated in the collinear DGLAP framework. }
\end{figure}

To conclude we turn our attention to the modification of 
the nuclear parton densities induced by the constraints from inclusive charged
hadron production pseudo-data in {\tt EPS09}. 
For the sake of clarity, 
we present in Fig.~\ref{fig:eta-eps-w} the plots for both $\eta=0$ (central) and $\eta=2$ (right), with the upper plots corresponding to the sea quark nuclear ratios and the lower ones to the nuclear gluon ratio.
 The valence quarks turn out to be almost unaffected (since they are already quite constrained by
nuclear DIS data).
The same effects occur for both rapidities: the sea density slightly decreases and the error band reduces a bit for $x<10^{-2}$. 
In the case of the gluon, the central values show an enhancement for $0.07<x<0.2$ while for $x<10^{-2}$ it is suppressed and the error band shrinks around $50\%$. 
While it is true that the displacement of the central curves is less pronounced in the forward region, they are nonetheless fully compatible within uncertainties, and compatible also with the variation from Fig.~\ref{fig:dy-sg}.
The substantial error reduction in the small-$x$ nuclear gluon ratio confirm that this
observable is potentially very important to be included in nPDF fits once the LHC data
becomes available.

\subsection{Hadroproduction in the CGC framework \label{sec:chargedcgc}}

Now we consider the scenario in which pseudo-data has been generated in the Color Glass
Condensate scenario (se e.g. the review~\cite{Albacete:2013tpa}), following the approach of Ref.~\cite{Albacete:2012xq}.
Therefore, as opposed to the previous case, the underlying theory 
for pseudo-data is independent from the one used to compute the {\tt EPS09} predictions,
which is always the perturbative DGLAP framework.
We shall consider again the central and forward rapidity regions, with $\eta=2$, noting that
CGC and DGLAP predictions are known to differ more in the latter case.
 While identified meson production data at central rapidity from RHIC~\cite{Adler:2006wg,Adams:2003qm,Adams:2006nd,Abelev:2009hx} has already been included in nuclear PDF fits~\cite{Eskola:2009uj,deFlorian:2011fp}, data from the forward region has not been included as it introduced a rather strong tension with DIS measurements \cite{Eskola:2008ca}, which might arise from non-linear effects. 
It is then interesting to include both regions in our study.

\begin{figure}[t]
\begin{center}
\vspace*{-0.6cm}
\epsfig{figure=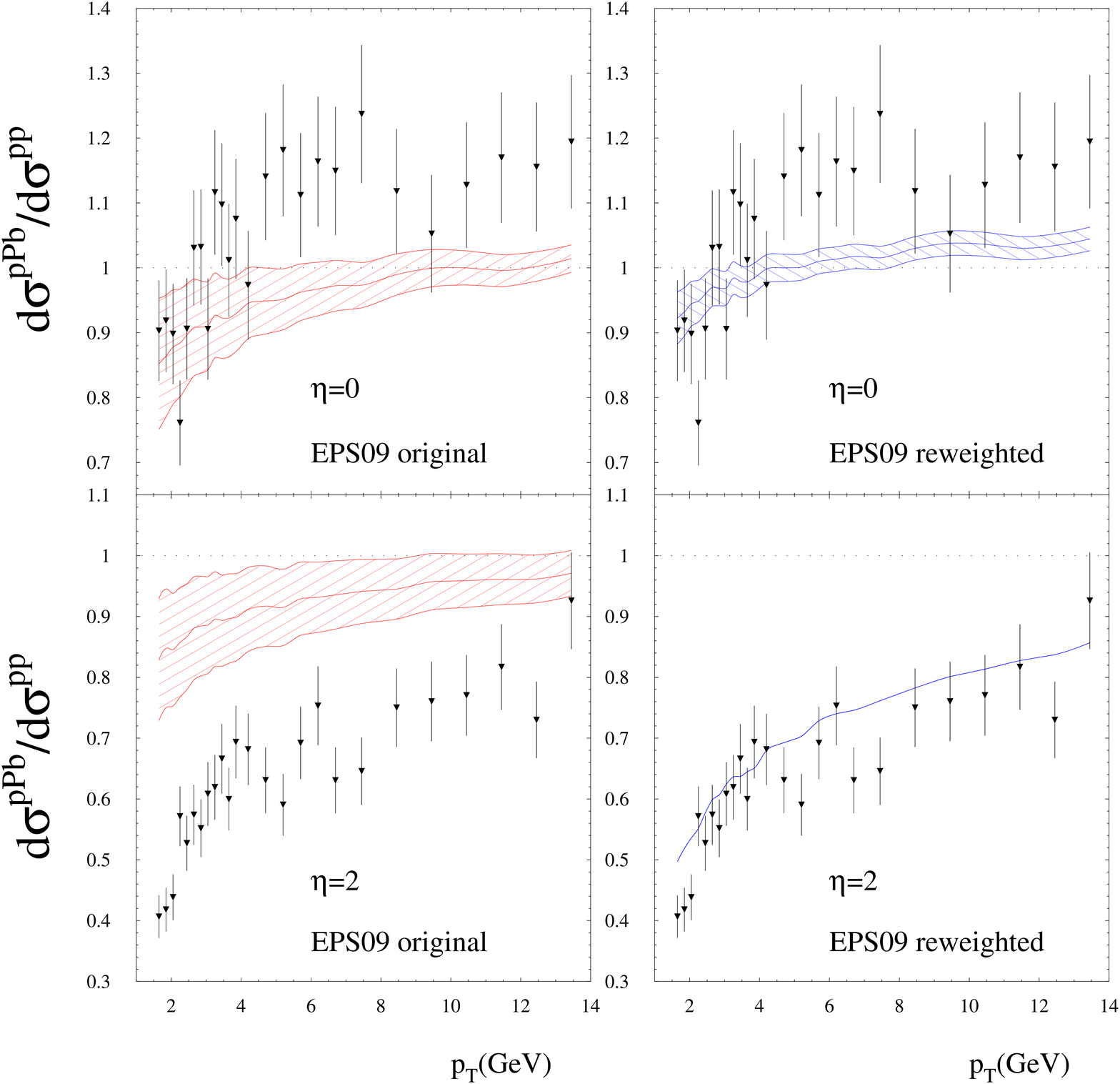,width=0.8\textwidth}
\end{center}
\vspace*{-0.5cm}
\caption{\label{fig:eta-cgc}
Upper plots: ratio of the charged hadron single inclusive cross section in $p$Pb with respect to pp $a$t central rapidity, as a function of the final hadron transverse momentum $p_T$ before (left) and after (right) the reweighting. 
Lower plots: the same for $\eta=2$. Note that the pseudo-data has been generated from the CGC predictions, while {\tt EPS09} predictions are based on the collinear
DGLAP framework.}
\end{figure}

To begin with, we present in Fig.~\ref{fig:eta-cgc} the nuclear cross section 
ratios corresponding to $\eta=0$ (upper plots) and $\eta=2$ (lower plots) before and after the reweighting, as a function of the transverse momentum $p_{T}$ of the final hadron. 
In Tab.~\ref{tab:chi2-cgc} we present the values for $\chi^{2}/n$ and $\langle \chi^{2}\rangle/n$ (with $n=25$) for both rapidities and the corresponding $N_{\rm eff}$.

Let us begin by discussing the impact of CGC pseudo-data from the central
rapidity region, $\eta = 0$.
Despite the fact that pseudo-data and theory predictions are generated with
different underlying theories, the agreement after reweighting is still reasonable (though clearly not as good as in the case of
DGLAP pseudo-data). 
The main impact of adding the pseudo-data is to
drag the central value upwards, as well as reducing the
PDF uncertainties. 
The effective number of replicas $N_{\rm eff}$t that  we obtain is $229$, and the final $\chi^{2}/n$ is $1.5$, indicating some tension between the CGC prediction and the DGLAP theory predictions that cannot be accomodated by a change in
shape or normalization of the
nuclear PDFs.

\begin{table}[h]
\begin{center}
\begin{tabular}{cccccccccc}
\hline
 & & $\eta=0$ & & \vline& & $\eta=2$     \\ \hline
 & $\chi^{2}/n$ & $\langle \chi^{2}\rangle/n$ & $N_{\rm eff}$ & \vline & $\chi^{2}/n$ & $\langle \chi^{2}\rangle/n$ & $N_{\rm eff}$  \\ \hline\hline
 Before & 2.25 & 2.76 & & \vline & 36.43 & 38.62 &      \\
 After & 1.50 & 1.58 & 229 & \vline& 1.85 & 1.85 & 1     \\  \hline
\end{tabular}
\caption{\label{tab:chi2-cgc} Same as Table~\ref{tab:chi2-eps} for the case
in which pseudo-data has been generated in the CGC framework.}
\end{center}
\end{table}

\begin{figure}[h]
\begin{center}
\vspace*{-0.6cm}
\epsfig{figure=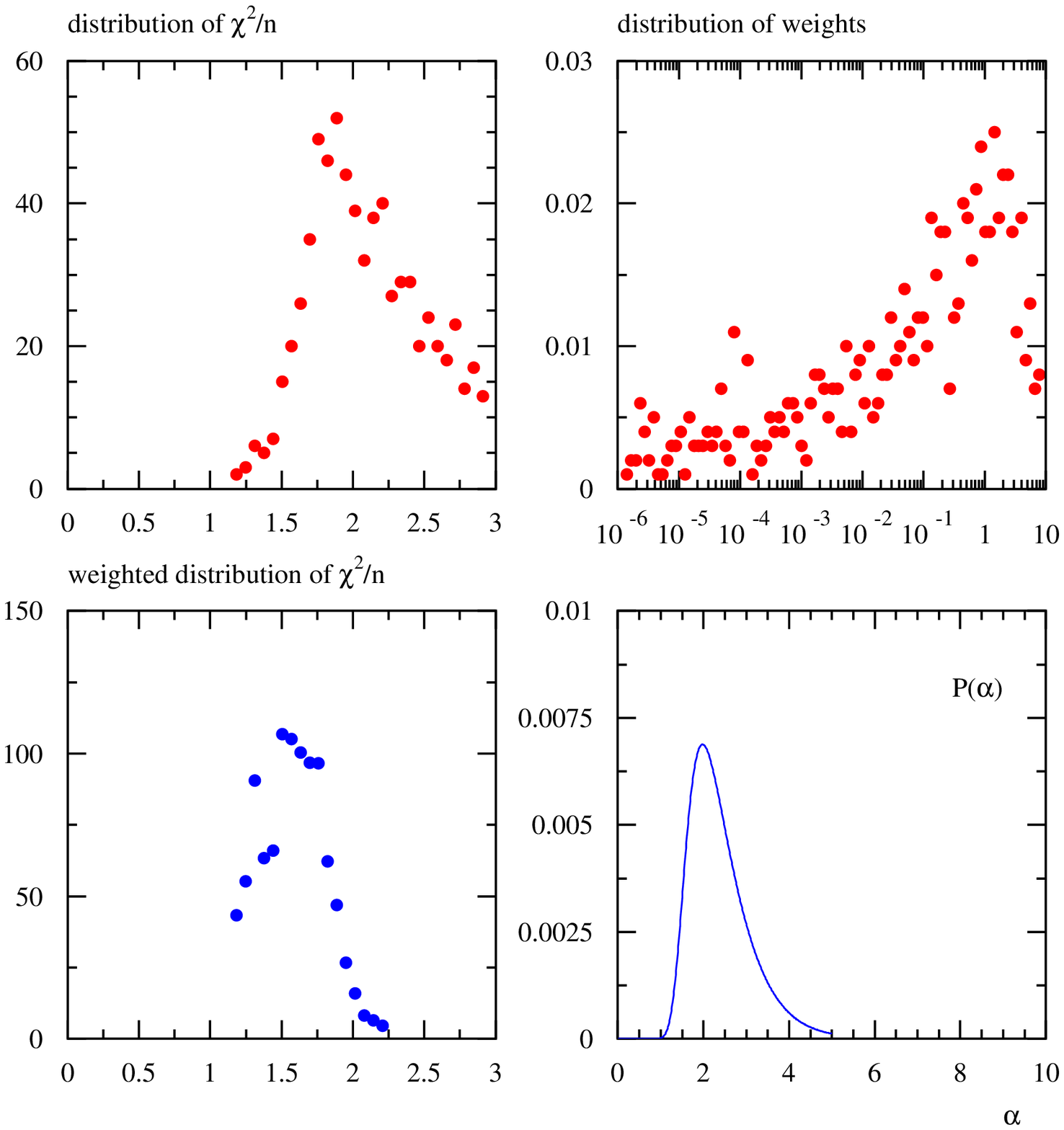,width=0.85\textwidth}
\end{center}
\vspace*{-0.5cm}
\caption{\label{fig:eta0-dist-cgc}
Same as Fig.~\ref{fig:dy-dist} for inclusive charged hadron central production
$\eta=0$, with pseudo-data generated in the CGC framework. 
}
\end{figure}

If we move to the consistency tests, shown in Fig.~\ref{fig:eta0-dist-cgc}, we find that the distribution of the $\chi^{2}/n$ is peaked around $2$ with a flat tail towards higher values, while after the reweighting the peak moves as expected towards lower values but remaining slightly above 1.
From the $\alpha$ distribution (lower right) we clearly see the inconsistency between data and theory predictions.
It implies that a satisfactory agreement would also be obtained at the expenses
of increasing experimental uncertainties by a factor two.
Conversely, the more precise the experimental data is, the more sensitive to
the differences between DGLAP and CGC predictions it will be, even in the
central region where these differences are moderate.

\begin{figure}[t]
\begin{center}
\vspace*{-0.6cm}
\epsfig{figure=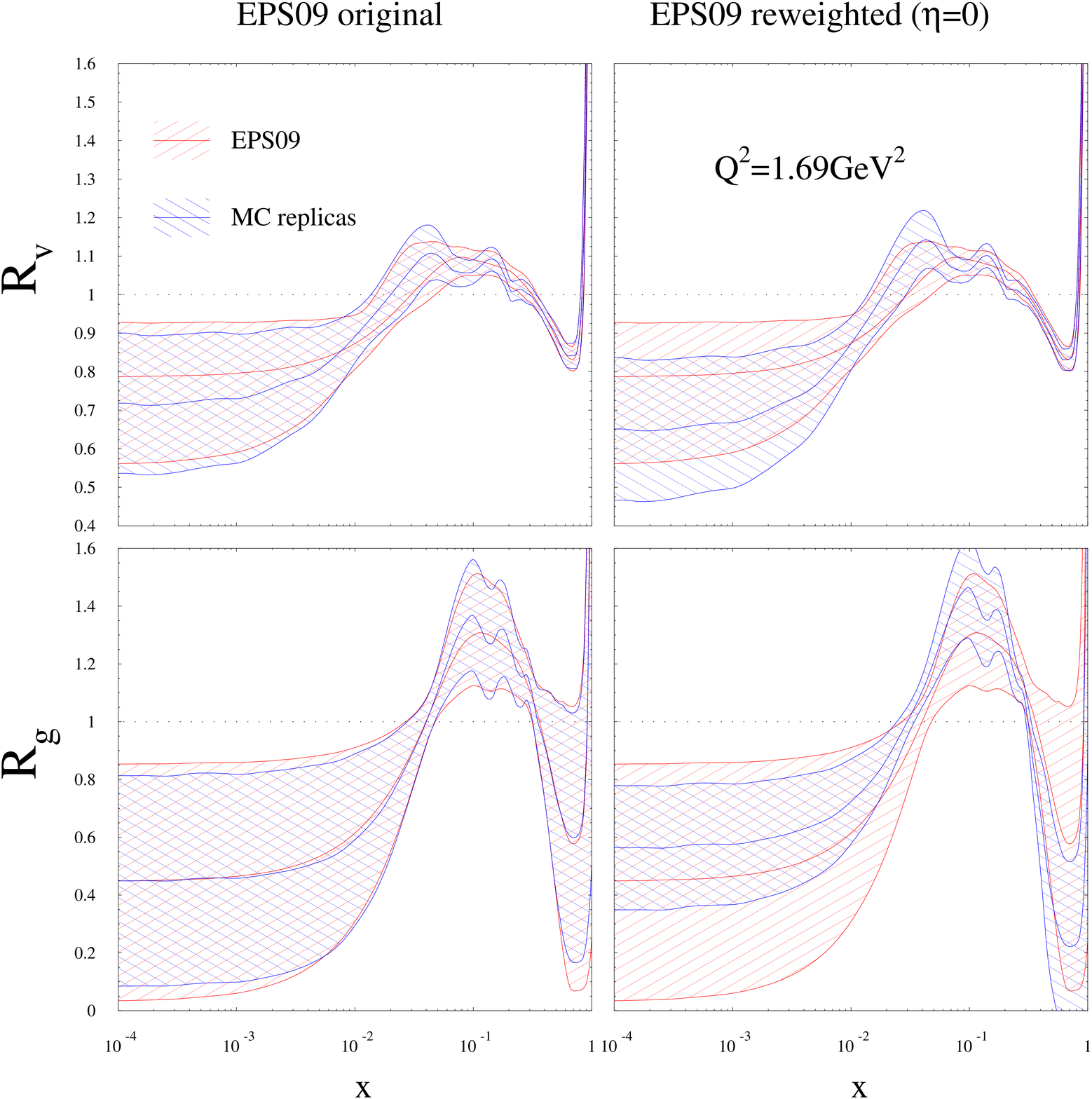,width=0.85\textwidth}
\end{center}
\vspace*{-0.5cm}
\caption{\label{fig:eta0-glu}
Same as Fig.~\ref{fig:dy-sg}, now for the nuclear modifications of up valence quarks (upper plots)
and gluons (lower plots) when $p$Pb LHC pseudo-data for inclusive charged hadron production
at central ($\eta=0$) rapidities is included in {\tt EPS09}. The
pseudo-data has been generated in the CGC framework. }
\end{figure}

As for the impact of the $\eta=0$ charged hadron production CGC pseudo-data in the nuclear PDF ratios, the valence distributions are affected, presenting the non-negligible decrease of the central value shown in the upper right plot of Fig.~\ref{fig:eta0-glu} for the up quark; no noticeable change is seen for the sea densities.
For the gluons instead (lower plots of Fig.~\ref{fig:eta0-glu}) the central value moves upwards and the uncertainty shrinks around 30\% for $x<10^{-2}$. 
Note that while the shriking of the gluon nPDF uncertainties is
 qualitatively similar as that seen in
Fig.~\ref{fig:eta-eps-w}, where pseudo-data was generated with DGLAP,
the trend of the central value is the opposite: in the CGC case,
we get a harder small-$x$ gluon, with a softer one in the case
of DGLAP pseudo-data.
This different trend can be explained because for central charged hadron
production, the CGC prediction leads to an enhancement as compared to
the DGLAP one.

Therefore we conclude that the data from the central region cannot
really discriminate between the two production scenarios, DGLAP and CGC.

Let us consider now the case in which CGC pseudo-data for charged
hadron production from the the forward region is included in the
nuclear PDF fit.
 From the reweighted cross section, shown in the lower right plot in Fig.~\ref{fig:eta-cgc}, we see that including the (inconsistent) pseudo-data into
the fit is not impossible, since the  inital $\chi^{2}/n$ of $36.4$ is
reduced down to around $1.8$ (see Table~\ref{tab:chi2-cgc}).
On the other hand, the effective number of replicas is only $N_{\rm eff}=1$,
implying that all MC replicas have been discarded except the one that
gives the best agreement with the CGC pseudo-data.
Under these situations of extreme incompatibilty, the PDF reweighting method
breaks down and becomes unreliable.

What this analysis suggests is that the differences
in forward charged hadron production in $p$Pb collisions at the LHC between
the DGLAP and CGC frameworks is so substantial that it is not possible
to absorb it by an update of the global nuclear PDF analysis, and
therefore that the potential for the discrimination between the two
scenarios is very good \footnote{Still, variations of the initial parameterisations of the nPDFs should be explored for this conclusion to be definitive.}.
Of course, measuring this very same processes at even forward rapidities
would make the differences even more striking.

\section{Summary and outlook}
\label{sec:summary}

The accurate determination of how parton distributions are modified in nuclei is an essential
input for our understanding of heavy ion collisions and of the physics of the quark-gluon plasma.
Current analyses of nuclear PDFs include all available experimental information on the partonic structure of the nucleus, with however the bulk of the data restricted to deep-inelastic scattering at medium and large $x$.
However, DIS is mostly sensitive to quark valence distribution, but fails to constrain sea and gluon densities, which
are most relevant at small-$x$. 
Drell-Yan data has been included since the first analyses to provide constraints to the sea quarks. These fixed target data is again limited to the intermediate region of $x$ and due to the large uncertainties is almost irrelevant e.g. for gluons. Recently, inclusive hadroproduction data from RHIC were also included in the fits to provide further constraints for the gluons. In any case, the amount of data is still quite limited both in kinematical coverage and in accuracy.

Therefore, the recent proton-lead run at the LHC has the potential to provide, before electron-ion colliders \cite{Accardi:2012hwp,AbelleiraFernandez:2012cc} become eventually available, very important
information on the nuclear modifications of PDFs, in particular for the small-$x$ gluon and
sea quarks which are now very badly constrained,  which in turn would improve our theoretical predictions for lead-lead collisions.
In addition, proton-lead data offers the possibility of uncovering new regimes of QCD, in particular,
non-linear effects such as those incorporated in the Color Glass Condensate scenario could be important enough to be disentangled from standard collinear factorisation.
Of course, the only way of quantifying the tension between the DGLAP and CGC frameworks is to perform
a new global nuclear fit and verify if the new data can be accommodated or not by modifications of the
nuclear PDFs in the collinear scenario.

With this motivation in mind, and 
since essentially no data in the hard-scattering regime is still available,
in this paper we have presented a first study of the potential of LHC proton-lead measurements
to constrain nuclear PDFs, based on simulated data.
From the methodological point of view, instead of performing new versions of the {\tt EPS09} fit,
 we have applied the technique of Bayesian PDF reweighting, which is now routinely used
in the case of proton PDFs.
We have considered two representative processes: Drell-Yan production and inclusive charged
particle production, both of which are sensitive to nuclear modifications of
both gluons and sea quarks.
For the case of Drell-Yan, we found that under conservative assumptions, available data
has the potential to reduce the PDF uncertainties on the small-$x$ nuclear sea quarks and
specially in the medium and small-$x$ nuclear gluon, where uncertainties can decrease by up
to a factor two.

Then we turned our attention to single inclusive hadroproduction.
 In this case two sets of pseudo-data were studied, one generated using the CGC scenario and other one generated using the same collinear DGLAP framework as the one used to produce the
theory predictions.
In the latter case, we find a similar impact as in the case of Drell-Yan pseudo-data, namely
reduction of nuclear PDF uncertainties in the sea quarks but specially on the gluon.
When pseudo-data is generated using CGC predictions, we find that the global nuclear fit is able to absorb the non-linear effects only for pseudo-data in the central region, 
while pseudo-data in the forward
region was manifestly incompatible with the DGLAP predictions.
Therefore, this process is interesting whatever the underlying behavior in real
data turns out to be:
if non linear effects are small, very useful constraints on virtually unknown nuclear
PDFs will be derived; if on the other hand they are large, it is likely that experimental accuracy
is enough to clearly identify the onset of the saturation dynamics.

Our study provides a first quantitative estimate of the potential of the proton-lead data to constrain PDFs,
and confirms that such experimental results should be an essential ingredient
of nuclear global PDF fits in the coming years, and thus become crucial for improving our understanding of heavy ion collisions.
Of course, we have used some simplified assumptions, and crude estimates of the experimental
uncertainties, so at the end only when the actual LHC measurements become available we will
be able to quantify the impact of the data on the nuclear PDFs.

From the methodological point of view, the availability of a Monte Carlo version of {\tt EPS09}
implies that the experimental groups themselves can study the impact of their data on nuclear PDF
by means of the reweighting method discussed here, without the need to wait for an updated fit.
With this motivation, {\tt EPS09MC} has been made publicly available.
The driver code, data files and documentation to use the {\tt EPS09} Monte Carlo sets (both the symmetric and asymmetric cases are provided) can be obtained
from
\begin{center}
\url{http://igfae.usc.es/hotlhc/index.php/software}.
\end{center}

\section*{Acknowledgments}
We thank Kari Eskola and Hannu Paukkunen for useful discussions, and 
 Javier Albacete for providing us with the CGC predictions for charged
particle production. 
This work is supported by European Research Council grant HotLHC ERC-2011-StG-279579; by Ministerio de Ciencia e Innovacion of Spain under projects FPA2008-01177, FPA2009-06867-E and FPA2011-22776; by Xunta de Galicia (Conselleria de Educacion and Conselleria de Innovacion e Industria - Programa Incite); by the Spanish Consolider-Ingenio 2010 Programme CPAN and by FEDER. 
P.~Z. thanks the hospitality of the CERN TH Unit where part of this
work was performed.
 J.~R. was partially supported by a Marie Curie Intra--European Fellowship of the European Community's 7th Framework Programme under contract number PIEF-GA-2010-272515.


\end{document}